\documentclass{article}

\usepackage{PRIMEarxiv}

\usepackage[utf8]{inputenc} % allow utf-8 input
\usepackage[T1]{fontenc}    % use 8-bit T1 fonts
\usepackage{hyperref}       % hyperlinks
\usepackage{url}            % simple URL typesetting
\usepackage{booktabs}       % professional-quality tables
\usepackage{amsfonts}       % blackboard math symbols
\usepackage{nicefrac}       % compact symbols for 1/2, etc.
\usepackage{microtype}      % microtypography
\usepackage{lipsum}
\usepackage{fancyhdr}       % header
\usepackage{graphicx}       % graphics
\graphicspath{{media/}}     % organize your images and other figures under media/ folder
\usepackage{verbatim}
\usepackage{amsmath}
\usepackage{array}
\usepackage[round]{natbib}
\newcolumntype{H}{>{\setbox0=\hbox\bgroup}c<{\egroup}@{}}
\title{Covariance matrix filtering and portfolio optimisation: the Average Oracle vs Non-Linear Shrinkage and all the variants of DCC-NLS}
\author{
  Christian Bongiorno and Damien Challet \\
Université Paris-Saclay\\CentraleSupélec\\
Laboratoire de Mathématiques et Informatique pour la Complexité et les Systèmes\\
91192 Gif-sur-Yvette, France\\
 \texttt{christian.bongiorno@centralesupelec.fr}\\
 \texttt{damien.challet@centralesupelec.fr}}

\begin{document}
\maketitle

\begin{abstract}

The Average Oracle, a simple and very fast covariance filtering method, is shown to yield superior Sharpe ratios than the current state-of-the-art (and complex) methods, Dynamic Conditional Covariance coupled to Non-Linear Shrinkage (DCC+NLS). We pit all the known variants of DCC+NLS (quadratic shrinkage, gross-leverage or turnover limitations, and factor-augmented NLS) against the Average Oracle in large-scale randomized experiments. We find generically that while some variants of DCC+NLS sometimes yield the lowest average realized volatility, albeit with a small improvement, their excessive gross leverage and investment concentration, and their 10-time larger turnover  contribute to smaller average portfolio returns, which mechanically result in smaller realized Sharpe ratios than the Average Oracle. We also provide simple analytical arguments about the origin of the advantage of the Average Oracle over NLS in a changing world.

%to compare the performance of the state-of-the-art.
%DCC-NLS is considered the state-of-art of covariance cleaning for portfolio optimization. Thousands of citations highlight such fame. However, a proper randomized set of experiments is missing. Compared with the most naive covariance filtering (a fixed set of eigenvalues independent from the input calibrated on an outdated dataset), DCC-NLS underperforms for all the standard metrics of portfolio optimization. 
\end{abstract}

% keywords can be removed

\section{Introduction}

Portfolio optimization is a problem of fundamental importance. It rests on determining and exploiting the dependency structure between asset price returns, some encoded in the covariance matrix in a first approximation. 
The main objective of portfolio optimization is to minimize the volatility while enhancing returns, thereby improving the Sharpe Ratio. Given the inherent difficulty in predicting returns, the emphasis is often placed on effectively minimizing variance, highlighting the essential role of a reliable covariance matrix estimation.  Focusing on reducing variance while monitoring other metrics such as realized profit, Sharpe Ratio, turnover, portfolio concentrations, and gross leverage is therefore crucial for mitigating the exposure to other sources of risk. 
In this context, a proper estimation of the covariance matrix turns out to be quite a challenging task. Recent progresses in covariance matrix filtering are nothing but remarkable \citep{bun2017cleaning}, culminating in the current state-of-the-art Dynamic Conditional Covariance coupled with Non-Linear Shrinkage (DCC+NLS) \citep{engle2019largecov}. While the DCC part takes care of the dynamics, NLS proposes a target matrix that is much less noisy than a direct estimation. It rests on a series of results that show how to replace the eigenvalues of a covariance matrix with filtered ones to minimize the distance between the NLS-estimated matrix and the true one. Various extensions and refinements of these methods have been proposed to improve some aspects of the original DCC+NLS method \citep{ledoit2022markowitz,ledoit2022quadratic}.

Yet, DCC+NLS rests on strong assumptions. NLS, in particular, applies if the long-term true covariance matrix is constant, while the DCC part takes care of temporal variations. Although this seems the ideal combination, an alternative approach consists of accounting only for the average unconditional shift in the covariances among consecutive time periods, which is the basic idea behind the filtering method known as the Average Oracle (AO hereafter)  \citep{bongiorno2021cleaning}: it uses a fixed set of eigenvalues independent from the input and calibrated on an outdated dataset.  Whereas this method seems to stand little chance against the sophistication of DCC+NLS, we show here that it consistently yields better portfolios than DCC+NLS, despite being orders of magnitude faster and simpler to implement. This raises substantial questions about the efficiency of complex methods in real scenarios and underscores the challenge of determining which information is relevant and preserved over time.%, suggesting a potential limitation of overly sophisticated approaches in the modeling of the dynamic nature of financial markets.

We compare various covariance filtering methods with randomized asset universes. A common approach consists in building portfolios with an investment universe made of the $n$ stocks with the largest capitalization either at the time of writing or historically, and a single portfolio is computed at a given time, leading to a single set of performance metrics. We believe that this approach can be improved in several respects:  high-capitalization stocks at a given date have, on average, outperformed during the past; second, having a single number for each metric does not allow us to measure confidence intervals without introducing assumptions such as normality or time invariance, both of them known to be false in real financial markets. 
To overcome these issues, we propose working with a universe composed of the top $N\gg 1$ capitalized stocks and then randomly sampling $n \ll N$ to build an investment universe. In addition, the universe of $N$ stocks must be selected at the start of the investment period, using only the information available at the time. If the capitalization of some of the stocks changes dramatically, such stocks should be sold. Since $n \ll N$, computations can be carried out over many realizations, and the confidence intervals can be determined in a non-parametric way via bootstraps. 

Our final contribution is to propose a simple covariance evolution framework that assumes that the true covariance matrix may change only at the end of each calibration period. This makes it easy to compare AO with NLS in a changing world and highlighs the fact that what matters are not only the eigenvalues but also the fact that they are weighted by the average eigenvector overlap between two successive covariance matrices. When the typical rotation between two such matrices is larger than the random rotation due to estimation noise, and provided that the eigenvalues do not change too much, AO beats NLS.

\section{Methods}
\subsection{Universe construction}\label{sec:univ}
To carry out our analysis, we outline here our data selection approach which is carefully crafted to mirror, as closely as possible, the actual way in which data selection is executed in real life, thus minimizing all possible selection biases.

Consider day $t$ an in-sample window $[t-\Delta t_{in}, t]$ and an out-of-sample window $]t, t+\Delta t_{out}]$. The selection of available assets will be the top $N$ most capitalized stocks of the US market on the day $t$ such that all the stocks are listed in $[t-\Delta t_{in},t+\Delta t_{out}]$; using the fact that the chosen assets are listed in the future involves a limited amount of foresightedness. In addition, we ask that the selected assets have less than $20\%$ zero or missing returns in $[t-\Delta t_{in}, t]$, and that any pair of assets should have a Pearson correlation coefficient smaller than $0.95$~\cite{de2021factor}.

\subsection{Covariance filtering methods}

The following list enumerates all the covariance estimation methods tested in this paper.
\begin{enumerate}
    \item \textbf{NotFilt1200} Sample covariance  with $\Delta t_{in}=1200$.
    \item \textbf{NotFilt240} Sample covariance  with $\Delta t_{in}=240$.   
    \item \textbf{DCC-QIS} with quadratic shrinkage \citep{ledoit2022markowitz}.
    \item \textbf{DCC-QuEST} DCC with the nonlinear shrinkage based on the QuEST function \citep{engle2019largecov}.
    \item \textbf{1AFM-DCC-QuEST} DCC with the nonlinear shrinkage based on the QuEST function with a factor taken from the Fama and French~\citep{de2021factor}.
    \item \textbf{1AFM-DCC-QIS} DCC with quadratic shrinkage with a factor taken from the Fama and French~\citep{de2021factor}.
    \item \textbf{NLS1200} nonlinear shrinkage based on the QuEST function  with $\Delta t_{in}=1200$ ~\citep{ledoit2017nonlinear}.
    \item \textbf{QIS1200} quadratic shrinkage with $\Delta t_{in}=1200$  \cite{ledoit2022quadratic}.
    \item \textbf{QuEST240} nonlinear shrinkage based on the QuEST function  with $\Delta t_{in}=240$ ~\citep{ledoit2017nonlinear}.
    \item \textbf{QIS240} quadratic shrinkage  with $\Delta t_{in}=240$ \cite{ledoit2022quadratic}.
    \item \textbf{AO1200} Average Oracle eigenvalues computed in the 1900-2000 period with $\Delta t_{in}=1200$ \citep{bongiorno2021cleaning}.    
    \item \textbf{AO240} Average Oracle eigenvalues calibrated in the 1900-2000 period with $\Delta t_{in}=240$ \citep{bongiorno2021cleaning}.
  %  \item \textbf{EQ} Equally weighted portfolio.
\end{enumerate}

DCC, NLS, and QIS were run with the original MatLab implementation available \citep{code:ledoit}.

\section{Empirical Analysis}
\subsection{Randomized experiments}\label{sec:setup}
According to \cite{engle2019largecov}, DCC performs better for $\Delta t_{in}=1200$ days; therefore, we use this calibration window. In this experiment, we fixed $n=100$ assets randomly sampled from the top $N=500$ stocks per capitalization at time $t$ as described in Sec.~\ref{sec:univ}. We tested both $\Delta t_{out}= 5$ and $20$. We performed a near yearly rebalancing of $240$ days, which consists of $48$ rebalancing for the $\Delta t_{out}= 5$ and $12$ rebalancing for $\Delta t_{out}= 20$. At each rebalancing, if some stocks of the portfolio exit the universe $N$ of the top $500$ capitalized stocks, such stocks will be sold and substituted with a random selection from the universe $N$ defined on the current day. The initial day $t$ is selected randomly in the range [2000-01-03,2021-01-20], which includes the 2008-2009  and 2020 COVID crises.

To measure the respective performance of all the above methods, we set the transaction cost to $5$ bps, which is considered a low transaction cost  \citep{ledoit2022markowitz}; larger transactions would increase the advantage of AO over DCCs. Transaction costs are paid only at each rebalancing and also account for the stock price drift in the portfolio.

We performed $10,000$ simulations by randomly sampling $t$ and the initial set of stocks $n$ available at time t. 
Tables~ \ref{tab:n100_dt5_LS} and \ref{tab:n100_dt20_LS} report  performance metrics of long-short portfolios, for weekly ($\Delta t=5$ days) and monthly ( $\Delta t=$20 days) rebalancing. First, there is no significant difference between QIS and QuEST-based methods for any performance measure; for this reason, we refer to them as DCC-QIS/QuEST. DCC-QIS/QuEST and AO240 achieve the smallest realized volatility, which is not statistically different even after 10,000 simulations. We note that 240 timesteps correspond to the timescale fitted by the DCC process. In other words, the complex machinery of DCC+NLS models the covariance matrix at a similar timescale as AO240. What is still remarkable is that AO240, despite its simplicity, leads to the same result. We also note that AO1200 has only marginally higher volatility (approximately $7\%$). 

There are large differences, however, in three key portfolio metrics. First, the Sharpe ratio of A01200 is $14\%$ ($\Delta t_{out}=20$) and $37\%$ ($\Delta t_{out}=5$) larger than that of DCC-QIS/QuEST, which is a large and significant difference. Interestingly, QIS/QuEST1200 has a systematically higher SR than their respective DCC versions (yet smaller than those of AO). In passing, even an unfiltered covariance matrix with 1200 days of calibration yields a larger Sharpe ratio, which is consistent with the well-known fact that when the number of $\Delta t_{in}$ is larger than about 10$n$, covariance filtering is not needed.  

Even more, the largest difference between AO and DCC-QIS/QuEST resides in the turnover. We report two kinds of turnover metrics: one defined by the change of weights between two rebalancing times prescribed by a given method, and one which includes, in addition, the change of asset prices, hence, weights, during the holding periods (denoted by Turnover+drift in all the tables). , which is 12-19 times larger for the latter. This comes from two main differences. First, the gross leverage is approximately 45\% larger for DCC-QIS/QuEST than AO1200. Second, the diversification of the portfolio from AO is almost twice that of DCC-QIS/QuEST.

The 1AFM improves a little bit the performance of DCC-QIS/QuEST because it gives more stable portfolios,  but at the cost of even larger gross leverage. 

%The same observations are valid for long-only portfolios, with some differences between the relative importance of turnover. One additionally notices that unfiltered covariance matrices with $\Delta t_{in}=1200$ yields undistinguishable results from QIS1200 and QuEST1200, which is not surprising as $n$ is much smaller than $\Delta t_{in}$, but AO1200 still does clearly better. What is surprising though, is that DCC-QIS/QuEST only brings small improvements as regards realized volatility. 

In summary, from the observations above, it seems that the DCC methods fail to give reliable (stable enough) estimates of the future covariance matrix and thus overbet on very concentrated portfolios that, unfortunately, change very quickly without capturing the relevant information, leading to poorer Sharpe ratios, higher gross-leverage, and low diversification.  

\begin{table}
    \centering
\begin{tabular}{llllllll}
\toprule
 & SR & MEAN & VOL & Turnover & Turnover+drift & GrossLev & $N_\textrm{eff}$ \\
\midrule
NotFilt1200 & 1.101 & 0.098 & 0.113 & 0.174 & 0.250 & 2.765 & 6.368 \\
NotFilt240 & 0.745 & 0.071 & 0.125 & 0.974 & 1.050 & 4.151 & 3.290 \\
AO1200 & \textbf{1.229}* & \textbf{0.112}* & 0.114 & \textbf{0.068}* & \textbf{0.119}* & \textbf{1.702}* & \textbf{13.981}* \\
AO240 & 1.167 & 0.095 & \textbf{0.105}* & 0.181 & 0.22 & 1.729 & 13.213 \\
DCC-QIS & 0.894 & 0.069 & 0.105* & 1.308 & 1.335 & 2.543 & 5.856 \\
DCC-QuEST & 0.895 & 0.069 & 0.105* & 1.307 & 1.333 & 2.542 & 5.852 \\
QIS1200 & 1.123 & 0.098 & 0.112 & 0.150 & 0.221 & 2.536 & 7.931 \\
QIS240 & 1.022 & 0.083 & 0.108 & 0.427 & 0.480 & 2.538 & 9.675 \\
QuEST1200 & 1.122 & 0.098 & 0.112 & 0.156 & 0.226 & 2.531 & 7.992 \\
QuEST240 & 1.027 & 0.084 & 0.108 & 0.402 & 0.456 & 2.512 & 10.011 \\
DCC-QIS-1F & 1.071 & 0.089 & 0.108 & 0.807 & 0.846 & 2.615 & 6.089 \\
DCC-QuEST-1F & 1.073 & 0.089 & 0.108 & 0.803 & 0.843 & 2.612 & 6.096 \\
\bottomrule
\end{tabular}

    \caption{$\Delta t_{out}=5$, $n=100$, average of $10,000$ simulations. GMV with long and short positions.  All the values not statistically distinct from the best value according to a bootstrap confidence interval at $95\%$ are marked in bold.}
    \label{tab:n100_dt5_LS}
\end{table}

\begin{table}
    \centering
\begin{tabular}{llllllll}
\toprule
 & SR & MEAN & VOL & Turnover & Turnover+drift & GrossLev & $N_\textrm{eff}$ \\
\midrule
NotFilt1200 & 1.111 & 0.101 & 0.116 & 0.414 & 0.639 & 2.763 & 6.376 \\
NotFilt240 & 0.811 & 0.079 & 0.13 & 2.008 & 2.197 & 4.158 & 3.305 \\
AO1200 & \textbf{1.230}* & \textbf{0.113}* & 0.117 & \textbf{0.173}* & \textbf{0.326}* & \textbf{1.705}* & \textbf{13.921}* \\
AO240 & 1.156 & 0.095 & \textbf{0.109}* & 0.422 & 0.540 & 1.733 & 13.218 \\
DCC-QIS & 1.07 & 0.088 & 0.110* & 2.082 & 2.117 & 2.544 & 5.962 \\
DCC-QuEST & 1.07 & 0.088 & 0.110* & 2.080 & 2.115 & 2.543 & 5.958 \\
QIS1200 & 1.132 & 0.101 & 0.115 & 0.358 & 0.569 & 2.535 & 7.938 \\
QIS240 & 1.039 & 0.086 & 0.112 & 0.891 & 1.040 & 2.545 & 9.718 \\
QuEST1200 & 1.132 & 0.101 & 0.115 & 0.358 & 0.568 & 2.531 & 8.000 \\
QuEST240 & 1.043 & 0.086 & 0.112 & 0.833 & 0.985 & 2.519 & 10.058 \\
DCC-QIS-1F & 1.152 & 0.099 & 0.112 & 1.279 & 1.389 & 2.624 & 6.085 \\
DCC-QuEST-1F & 1.153 & 0.100 & 0.112 & 1.275 & 1.385 & 2.622 & 6.091 \\
\bottomrule
\end{tabular}

\caption{$\Delta t_{out}=20$, $n=100$, randomized asset universe, average of $10,000$ simulations. GMV with long and short positions. All the values not statistically distinct from the best value according to a bootstrap confidence interval at $95\%$ are marked in bold. }
    \label{tab:n100_dt20_LS}
\end{table}

%\begin{comment}

\begin{table}
    \centering
\begin{tabular}{lllllll}
\toprule
 & SR & MEAN & VOL & Turnover & Turnover+drift & $N_\textrm{eff}$ \\
\midrule
NotFilt1200 & 1.134 & 0.108 & 0.124 & 0.044 & 0.07 & 10.362 \\
NotFilt240 & 1.072 & 0.092 & 0.116 & 0.149 & 0.169 & 9.294 \\
AO1200 & \textbf{1.187}* & \textbf{0.114}* & 0.125 & \textbf{0.037}* & \textbf{0.064}* & \textbf{16.676}* \\
AO240 & 1.122 & 0.098 & 0.116 & 0.106 & 0.127 & 15.528 \\
DCC-QIS & 1.041 & 0.084 & \textbf{0.110}* & 0.597 & 0.605 & 8.029 \\
DCC-QuEST & 1.040 & 0.084 & \textbf{0.110} & 0.597 & 0.606 & 7.995 \\
QIS1200 & 1.133 & 0.107 & 0.124 & 0.043 & 0.069 & 11.694 \\
QuEST1200 & 1.133 & 0.107 & 0.124 & 0.044 & 0.07 & 11.728 \\
QIS240 & 1.079 & 0.093 & 0.116 & 0.126 & 0.146 & 14.719 \\
QuEST240 & 1.076 & 0.093 & 0.116 & 0.127 & 0.148 & 14.588 \\
DCC-QIS-1F & 1.094 & 0.100 & 0.121 & 0.306 & 0.32 & 8.570 \\
DCC-QuEST-1F & 1.094 & 0.100 & 0.121 & 0.306 & 0.321 & 8.566 \\
\hline
EQ & 0.802 & 0.092 & 0.182 & 0.008 & 0.054 & 100. \\
\bottomrule
\end{tabular}
    \caption{$\Delta t_{out}=5$, $n=100$, randomized asset universe, average of $10,000$ simulations. GMV with long-only positions.}
    \label{tab:n100_dt5_Long}
\end{table}

\begin{table}
    \centering
\begin{tabular}{lllllll}
\toprule
 & SR & MEAN & VOL & Turnover & Turnover+drift & $N_\textrm{eff}$ \\
\midrule
NotFilt1200 & 1.124 & 0.108 & 0.127 & 0.115 & 0.197 & 10.346 \\
NotFilt240 & 1.038 & 0.09 & 0.12 & 0.344 & 0.399 & 9.322 \\
AO1200 & \textbf{1.177}* & \textbf{0.114}* & 0.127 & \textbf{0.097}* & \textbf{0.182}* & \textbf{16.611}* \\
AO240 & 1.095 & 0.096 & 0.12 & 0.253 & 0.317 & 15.535 \\
DCC-QIS & 1.074 & 0.091 & \textbf{0.116}* & 0.926 & 0.931 & 8.130 \\
DCC-QuEST & 1.073 & 0.091 & 0.116* & 0.927 & 0.932 & 8.096 \\
QIS1200 & 1.122 & 0.107 & 0.126 & 0.111 & 0.194 & 11.679 \\
QuEST1200 & 1.122 & 0.107 & 0.126 & 0.112 & 0.194 & 11.714 \\
QIS240 & 1.049 & 0.090 & 0.120 & 0.289 & 0.350 & 14.773 \\
QuEST240 & 1.046 & 0.090 & 0.120 & 0.288 & 0.349 & 14.646 \\
DCC-QIS-1F & 1.103 & 0.102 & 0.124 & 0.469 & 0.513 & 8.582 \\
DCC-QuEST-1F & 1.102 & 0.102 & 0.124 & 0.469 & 0.513 & 8.577 \\
\hline
EQ & 0.793 & 0.089 & 0.183 & 0.0288 & 0.156 & 100 \\
\bottomrule
\end{tabular}
    \caption{$\Delta t_{out}=20$, $n=100$, randomized asset universe, average of $10,000$ simulations. GMV with long-only positions.}
    \label{tab:n100_dt20_Long}
\end{table}

\subsection{Large Portfolios}
\cite{engle2019largecov} suggest that the DCC+NLS method performs better for large $n$. However, randomized experiments with $n=1000$  would require a much larger stock universe (say, $N=10,000$). But this is problematic for two additional reasons: first, DCC and NLS are very slow when $n$ is large; second, assets with a small capitalization are not traded every day, which sometimes prevents a proper calibration of GARCH models.

\begin{figure}
    \centering
    \includegraphics[width=0.3\columnwidth]{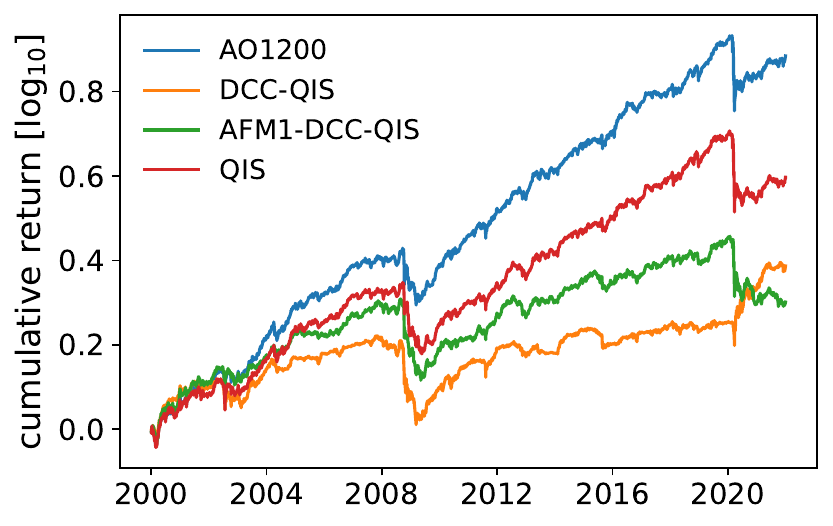}
    \includegraphics[width=0.3\columnwidth]{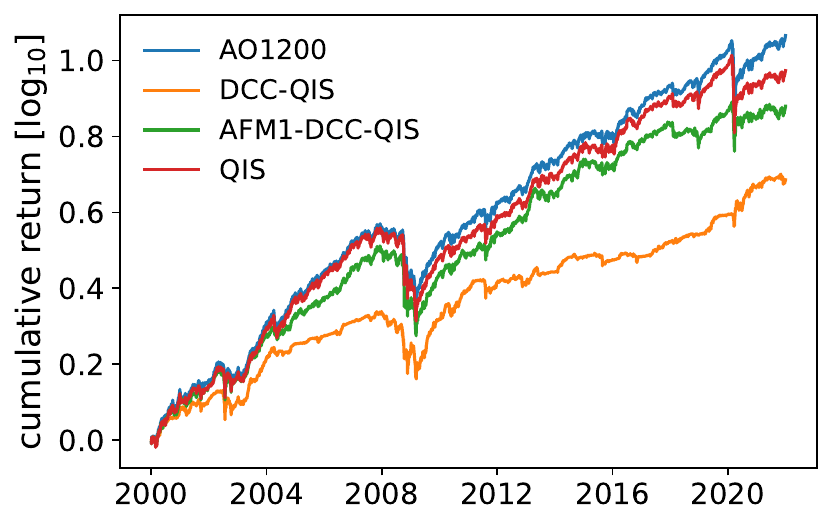}
    \caption{Large portfolio case: $n=1000$. Left plot: long-short portfolios, right plot: long-only portfolios. Portfolio rebalancing every week.}
    \label{fig:large_ptf}
\end{figure}

Therefore, we test a more standard approach. We performed two simulations with $\Delta t_{out}= 5$  using only the top $n=1000$ most capitalized stocks at each rebalancing time. Likewise, in the previous case, if a stock exits the top $n=1000$ most capitalized universe, it is replaced by a new suitable asset. The simulations start on 2000-01-03 and end on 2021-12-30.

\begin{table}
    \centering
\begin{tabular}{llllllll}
\toprule
 & SR & MEAN & VOL & Turnover & Turnover+drift & GrossLev & $N_\textrm{eff}$ \\
\midrule
NotFilt1200 & -0.6 & -0.097 & 0.162 & 7.235 & 7.282 & 17.562 & 2.013 \\
AO1200 & \textbf{1.009}* & \textbf{0.093}* & 0.093 & \textbf{0.210} & \textbf{0.242} & 2.961* & \textbf{46.147} \\
QIS & 0.689 & 0.063 & 0.092 & 0.729 & 0.761 & 4.859 & 29.179 \\
DCC-QIS & 0.484 & 0.041 & 0.084* & 2.238 & 2.243 & 3.823 & 12.957 \\
AFM1-DCC-QIS & 0.37 & 0.032 & 0.088 & 2.558 & 2.572 & 6.375 & 11.452 \\
\hline
AFM1-DCC-QIS-turn & 0.812 & 0.072 & 0.089 & 0.433 & 0.279 & 6.408 & 11.704 \\
DCC-QIS-turn & 0.972 & 0.078 & \textbf{0.080} & 0.405 & 0.296 & 4.800 & 10.851 \\
QIS-turn & 0.804 & 0.074 & 0.092 & 0.354 & 0.294 & 4.809 & 29.575 \\
DCC-QIS-gross & 0.528 & 0.044 & 0.083 & 1.929 & 1.933 & \textbf{2.862} & 13.494 \\
AFM1-DCC-QIS-gross & 0.556 & 0.048 & 0.087 & 1.585 & 1.591 & 2.961 & 14.199 \\
QIS-gross & 0.752 & 0.07 & 0.092 & 0.416 & 0.438 & 2.961 & 37.43 \\
\bottomrule
\end{tabular}
    \caption{Large long-short portfolio case ($n=1000$), single realization [2000-2021], the asterisk marks the best unconstrained method. }
    \label{tab:large_ptf_LS}
\end{table}

\begin{figure}
    \centering
    \includegraphics[width=0.3\columnwidth]{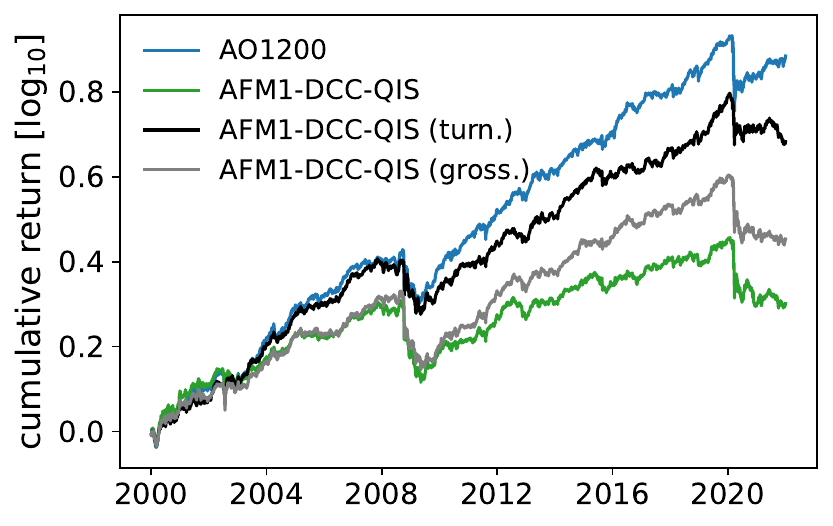}
    \includegraphics[width=0.3\columnwidth]{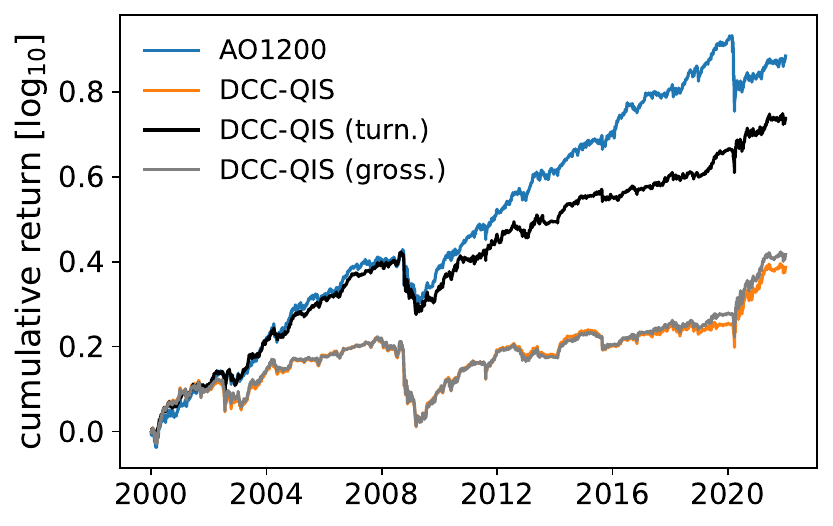}
    \includegraphics[width=0.3\columnwidth]{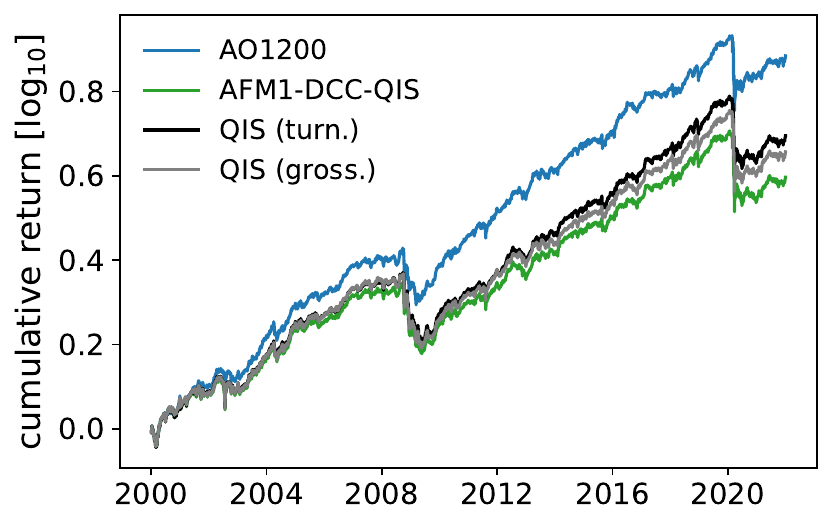}
    
    \includegraphics[width=0.3\columnwidth]{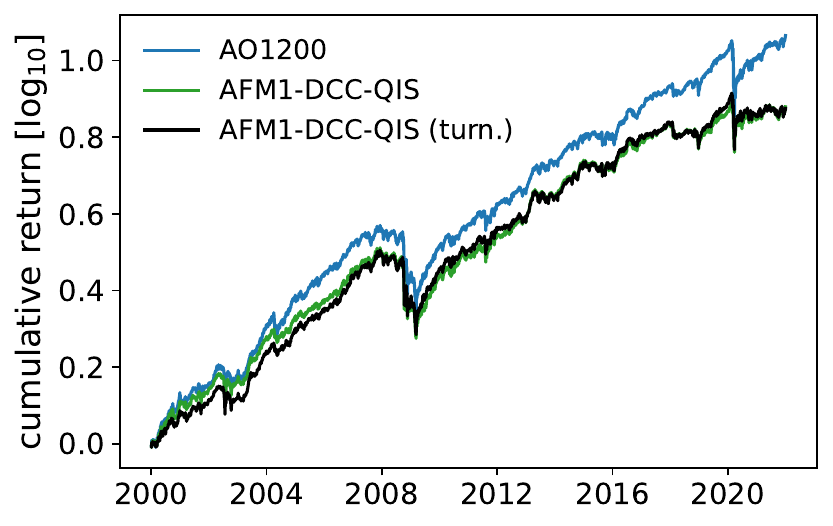}
    \includegraphics[width=0.3\columnwidth]{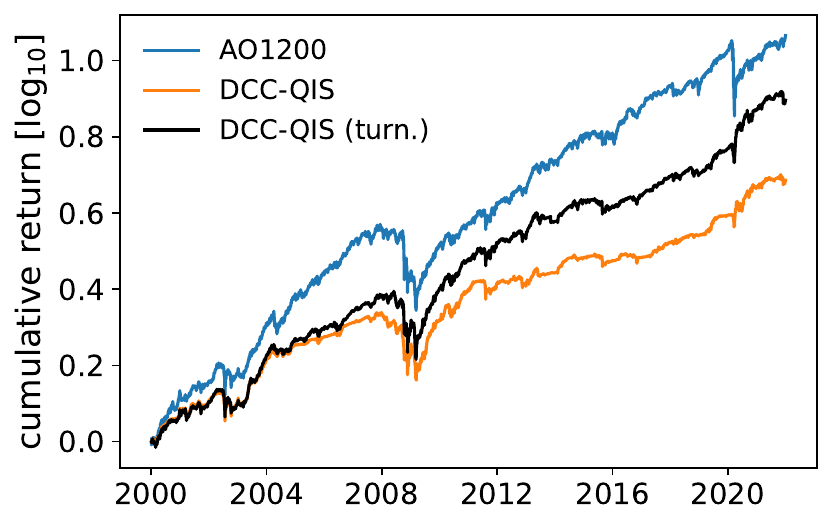}
    \includegraphics[width=0.3\columnwidth]{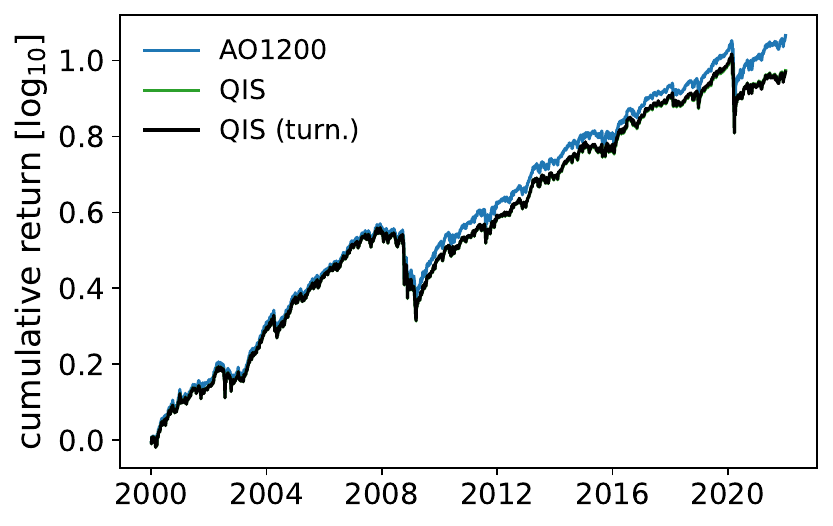}

    \caption{Large portfolio case ($n=1000$). Upper plots:  long-short portfolios, lower plots: long-only portfolios. Portfolio rebalancing every week}
    \label{tab:large_ptf_constr_LS}
\end{figure}

Figure \ref{fig:large_ptf}  reports the results of these experiments. While the superiority of AO in this setting is visually obvious, particularly regarding long-short portfolios,  Tables \ref{tab:large_ptf_constr_LS} and \ref{tab:large_ptf_constr_Long} give a more detailed breakdown: for long-short portfolios, AO has a much better Sharpe ratio than QIS, DCC-QIS and AFM1-DCC-QIS (107\%, 121.9\%, and respectively). This comes both from larger average returns and smaller realized volatility. In addition, as in the random universe experiment, the turnover of AO is approximately smaller by a factor of at least 10 than all the other methods, while having about half the gross leverage. Finally, the effective number of assets of AO is more than 20 times larger than those of the other methods.

The advantage of AO for long-only portfolios (Table \ref{tab:large_ptf_constr_Long}) is less impressive but still outperforms the other three methods in all metrics: its Sharpe ratio is significantly larger, once again because of both better average returns and realized volatility (except for DCC-QIS). AO has the same realized volatility as QIS and not filtered sample covariance (NotFilt1200). The turnover is also much smaller for AO, and the effective number of assets is much larger than all the other methods.

\subsection{Reducing Gross Leverage and Turnover for DCCs}

One of the major problems we observed with DCCs is their very large turnover and gross leverage. In the presence of transaction costs, this problem mechanically reduces the average return. To give all possible chances to DCC+NLS,  we now constrain the DCC methods to have a turnover or gross leverage bounded by those of AO1200 (we can only control the turnover in the case of long-only portfolios). We accounted for the fact that sometimes stocks exit the top 1000 and must be sold; accordingly, we only constrain the rebalancing that concerns stocks still in the top 1000 at the time of the rebalancing. 

\begin{table}
    \centering
\begin{tabular}{lllllll}
\toprule
 & SR & MEAN & VOL & Turnover & Turnover+drift & $N_\textrm{eff}$ \\
\midrule
NotFilt1200 & 0.772 & 0.095 & 0.124 & 0.117 & 0.122 & 18.142 \\
AO1200 & 0.897* & \textbf{0.112} & 0.125 & \textbf{0.083} & 0.089* & \textbf{45.559} \\
QIS & 0.832 & 0.103 & 0.124 & 0.1 & 0.105 & 34.838 \\
DCC-QIS & 0.869 & 0.072 & \textbf{0.083} & 0.635 & 0.636 & 6.765 \\
AFM1-DCC-QIS & 0.801 & 0.093 & 0.116 & 0.473 & 0.474 & 12.402 \\
\hline
AFM1-DCC-QIS-turn & 0.777 & 0.092 & 0.119 & 0.1 & 0.089 & 14.816 \\
DCC-QIS-turn & \textbf{0.980} & 0.094 & 0.096 & 0.134 & 0.123 & 14.47 \\
QIS-turn & 0.826 & 0.102 & 0.124 & 0.087 & \textbf{0.088} & 34.651 \\
%AO240 & 0.859 & 0.097 & 0.113 & 0.164 & 0.167 & 40.629 \\
\bottomrule
\end{tabular}
    \caption{Large long-only portfolio case ($n=1000$), single realization [2000-2021], the asterisk marks the best unconstrained method.}
    \label{tab:large_ptf_constr_Long}
\end{table}

While the unconstrained AO still beats all the DCC+NLS and NLS methods for long-short portfolios for all metrics except realized volatility, DCC-QIS, once constrained by AO1200, does bring the best Sharpe ratio, owing to its smaller realized volatility. 

%\end{comment}

\section{Average Oracle vs NLS in a simple dynamical model}

The results above raise the question of why the Average Oracle performs better than non-linear shrinkage when the true covariance matrix is time-dependent. This section is devoted to capture the essence of the difference between what the Average Oracle computes and the assumptions of the non-linear shrinkage estimator.

Let us assume that one needs to estimate covariance matrix $C$ between $N$ time series a time window of length $T$. The problem with many real-life systems is that the true covariance matrix $C$  evolves as a function of time. Thus, over the calibration window of length $T$, the true covariance matrix may not be constant, which complexifies both its estimation and mathematical computations.

Instead of assuming that $C$ evolves slowly as a function of time as DCC+NLS does, we assume that it changes at fixed times $kT$, $k\in {1,2,\cdots}$, which also are opportunistically assumed to correspond to the end of successive estimation windows. Equivalently,  time is split into slices indexed by $k$ and of length $T$: slice $k=[kT,(k+1)T[$. In other words, we assume that the true correlation matrix $C$ is constant during a time slice and may only switch between $S$ several possible values at the end of each time slice, possibly in a probabilistic way. This is a great simplification that makes it possible to derive simple bounds. The $S>1$ different covariance matrices are denoted by $C_s$, $s=1,\cdots,S$. During time slice $k$, the true covariance matrix index is $s(k)$, the former being denoted by  $C_{s(k)}$.

Assume that the current covariance matrix is $i=s(k)$. The spectral decomposition theorem states that $C_i=V_i\Lambda_i {V_i}^\dagger$, where the $\dagger$ symbol denotes the transpose operation, $V_i$  is the eigenvector matrix of $C_i$, and $\Lambda_i$ is the diagonal matrix of the eigenvalues of matrix $i$. We will also denote by $\lambda_i$ the vector of eigenvalues of $C_i$.

At end of the current period, the probability transition from true covariance matrix with index $i=s(k)$ to the one with index $j=s(k+1)$ is denoted by  $W_{i\to j}$ and the stationary frequency of state $i$ $P_i=\frac{1}{N}\sum_{j=1}^S W^\infty _{j\to i}$.  The eigenvectors of $C_j$ are given by a rotation of those of $C_i$: the rotation matrix $R_{ij}$ is defined as $R_{i\to j}=V_j {V_i}^\dagger$, so that $V_j=R_{i\to j}V_i$ and $V_i=R^{-1}_{i\to j}V_j$.

Estimating $C_i$ yields $\hat{V}_i$ and $\hat{\Lambda_i}$, noisy estimates  of 
$V_i$  and $\Lambda_i$. Computing the joint distribution of $\hat{V}_i$ and $\hat{\Lambda_i}$ is in general a hard problem (that can be tackled by sophisticated mathematical methods \cite{bun2017cleaning}).

By definition, the AO eigenvalues will be the average of Oracle eigenvalues computed from the estimated eigenvalues of $C_{s(k+1)}$ with the estimated eigenvectors of $C_{s(k)}$ over all possible pairs of correlation matrices, weighted by the transition probability, i.e.,

\begin{equation}\label{eq:AO_Wij}
E(\Lambda^{AO})=\sum_{ij}^SE[\hat P_i\hat W_{i\to j}\textrm{diag}(\hat{V}_i^\dagger\hat{C}_{j}\hat{V}_i)].
\end{equation}

A welcome simplification comes from the relationship
$$\textrm{diag}(\hat{V}_i^\dagger \hat{C}_{j}\hat{V}_i)=[(\hat{V}_i^\dagger\hat{V}_j)^{\circ2} \hat \lambda_j]=(\hat R_{i\to j})^{\circ 2}\hat\lambda_j
$$
where $X\circ Y$ is the element-by-element Hadamar product.
\cite{bongiorno2021cleaning} finds empirically that $$E[(\hat R_{i\to j})^{\circ2}\hat \lambda_j)]\simeq E[(\hat R_{i\to j})^{\circ2}]\ E[\hat \lambda_j]$$
to a very high accurancy, that is,
that the eigenvector overlap and the eigenvalue estimates are independent. Finally, Eq.\ \eqref{eq:AO_Wij} becomes

\begin{equation}\label{eq:AO_Wij_simpl}
E(\Lambda^{AO})\simeq\sum_{ij}^SE[\hat P_i\hat W_{i\to j}]E[(\hat R_{i\to j})^{\circ2}]\ E[\hat \lambda_j],
\end{equation}
as the transition process between the true covariance matrices is independent from estimation noises.

AO eigenvalues must be computed over a long calibration interval which should include as many transitions as needed to have reliable estimates at least of  the eigenvalues $\lambda_j$. When this calibration interval is infinitely long, 
\begin{equation}\label{eq:AO_Wij_simpl_inf}
E(\Lambda^{AO})\to \sum_{ij}^SP_i W_{i\to j}E[( R_{i\to j})^{\circ2}]\ \lambda_j.    
\end{equation}

Eqs \eqref{eq:AO_Wij_simpl} and \eqref{eq:AO_Wij_simpl_inf} make it clear that some cases simplify much the discussion: for example, when all the covariance matrix have the same eigenvalues, or if the overlap term does not depend on $i$ and $j$.

The simplest case (and the worst one for NLS) is a world in which $C_{s(k+1)}\ne C_{s(k)}$. For $S=2$, $s(k)=1+k\,\textrm{MOD}\, 2$, the world is cyclic and the AO eigenvalues are given by
$$
E(\Lambda^{AO})=E\left[(\hat R)^{\circ 2}\frac{\hat \lambda_1+\hat \lambda_2}{2}\right]\to E[( R)^{\circ 2}]\cdot \frac{ \lambda_1+ \lambda_2}{2}.
$$

On the contrary, NLS only uses information from the last time slice: when the true covariance matrix is $C_i$ during time slice $k$, its eigenvalues are given by
\begin{equation}\label{eq:Lambda_NLS}
    \hat \Lambda_{i}^{NLS}\to (\hat V_i^\dagger V_i)^{\circ 2}\hat \lambda_i
\end{equation}
in the limit of large $n$ and $T$ at fixed $n/T$ ratio. These eigenvalues minimize the Frobenius distance (average element-wise squares) $||\hat V_i \Lambda_{i}^{NLS}\hat V_i^\dagger-C_i||_2$. The magic of NLS is that it gives a closed formula for the eigenvalues that does not depend on the true matrix. It should be noted that $(\hat V_i^\dagger V_i)^{\circ 2}$ corresponds to the overlap matrix between the real eigenvectors and its estimates. Note as well that $\lambda_i$ is a noisier estimate for NLS than for AO in this context because it is performed over a single time slice.

The point is that the realized covariance matrix is different from $C_i$ in a non-constant world and thus AO incorporates both the average eigenvalues and the average overlap of the eigenvectors between the calibration and test time windows. In the simple cyclic word defined above, half of the terms of the AO eigenvalues do account for the future transition: one has
\begin{align}
||\hat V_1 \Lambda^{AO}\hat V_1^\dagger-C_2||_2&=||\hat V_1 E[( R)^{\circ 2}] \frac{ \lambda_1}{2}\hat V_1^\dagger+ V_1 E[( R)^{\circ 2}] \frac{\lambda_2}{2}\hat V_1^\dagger-C_2||_2
\end{align}
Now, 
$ E[( R)^{\circ 2}] \lambda_2$  are the Oracle eigenvalues when the true covariance matrix is $C_1$ in the estimation time slice and $C_2$  in the test time slice, hence are Frobenius optimal. Rearranging the terms, one has
\begin{align}
||\hat V_1 \Lambda^{AO}\hat V_1^\dagger-C_2||_2&=
||\frac{1}{2}\left(\hat V_1 E[( R)^{\circ 2}] \lambda_1\hat V_1^\dagger-C_2\right)+\frac{1}{2}\left(\hat V_1 E[( R)^{\circ 2}] \lambda_2\hat V_1^\dagger-C_2\right)||_2\\
&\le||\hat V_1 E[( R)^{\circ 2}] \lambda_1\hat V_1^\dagger-C_2||_2 
\end{align}
Finally 
$$
||\hat V_1 E[( R)^{\circ 2}] \lambda_1\hat V_1^\dagger-C_2||_2 \le ||\hat V_1 E[ (\hat V_i^\dagger V_i)^{\circ 2}] \lambda_1\hat V_1^\dagger-C_2||_2
$$
 as $E[ (\hat V_i ^\dagger V_i)^{\circ 2}]$ is a random rotation $\widehat{\delta R}$ due to eigenvector estimation noise, and does not bring $V_1$ closer to $V_2$ while $E[(\hat R)^{\circ2}]$ is the composition of the true rotation from $V_1$ to $V_2$ followed by a random rotation $\widehat{\delta R'}$ of the same order of magnitude as $\widehat{\delta R}$. As a consequence
\begin{align}
    ||C_1^{AO}-C_2||_2&\le||C_1^{NLS}-C_2||_2\\
    ||C_2^{AO}-C_1||_2&\le||C_2^{NLS}-C_1||_2.
\end{align}

The above result only holds when the true covariance matrix systematically changes at the end of each time slice. It is easy to generalize it to a cyclic world with $S$ states. When there is a finite probability that the true covariance matrix stays constant, there are cases when NLS performs better than AO, even in this simplifying context. Finally, the more generic case where the true covariance matrix may change slightly at each time step is currently out of reach of analytical methods.

\section{Conclusion}

Large-scale randomized experiments for mid-size portfolios (100 assets) allowed us to clarify the merits of a whole range of covariance filtering methods. Whereas the Average Oracle is a somewhat naive and very fast method not derived from an econometric framework, it still outperforms all the econometric methods tested here on four key portfolio metrics: Sharpe ratio, turnover, gross leverage, and diversification. DCC-QIS/QuEST yields a small improvement in realized volatility in a few cases.

%We also investigated the large portfolio case, reputed to bring out the real qualities of DCC-QIS/QuEST. While the picture did not change much regarding turnover, gross leverage, and diversification, realized volatility is remarkably better with unconstrained DCC-QIS/QuEST but only for long-only portfolios. Sharpe ratios are still (slightly) better with the Average Oracle, unless one constrains the turnover of DCC-QIS, which then loses much of its realized vo latility advantage.

The contrast between the simplicity of the Average Oracle and the sophistication of DCC-QIS/QuEST begs to incorporate ingredients from econometric methods, i.e., to leverage the qualities of both approaches in order to yield even better covariance filtering tools. This is left for future work.

\section*{Acknowledgments}

This publication used HPC resources from the ``M\'esocentre'' computing center of CentraleSup\'elec and \'Ecole Normale Sup\'erieure Paris-Saclay supported by CNRS and R\'egion \^{I}le-de-France.

%Bibliography
\bibliographystyle{unsrtnat}  
\bibliography{references}

\end{document}